\documentclass[preprint,showpacs,preprintnumbers,amsmath,amssymb,endfloats*]{revtex4}
\usepackage{graphicx}

\begin{document}
\def \ee {\varepsilon}
\thispagestyle{empty}
\title{
Comment on ``Precision measurement of the Casimir-Lifshitz
force in a fluid''
}

\author{B.~Geyer${}^1$,
G.~L.~Klimchitskaya${}^{1,2}$, U.~Mohideen${}^{3}$
and V.~M.~Mostepanenko,${}^{1,4}$
}

\affiliation{
${}^1$Center of Theoretical Studies and Institute for Theoretical 
Physics, Leipzig University,
D-04009, Leipzig, Germany\\
${}^2$North-West Technical University, Millionnaya Street 5,
St.Petersburg, 191065, Russia\\
${}^{3}$Department of Physics and Astronomy, University of 
California, Riverside, California 92521, USA \\
${}^4$Noncommercial Partnership  ``Scientific Instruments'', \\
Tverskaya Street 11, Moscow, 103905, Russia 
}

\begin{abstract}
Recently J.~N.~Munday and F.~Capasso 
[Phys. Rev. A {\bf 75}, 060102(R) (2007)] claimed that they have performed
a precision measurement of the Casimir force between a sphere and a
plate coated with Au, both immersed in ethanol. The measurement results
were claimed to be consistent with the Lifshitz theory. We demonstrate
that the calculation of the Casimir force between the smooth surfaces 
of the test bodies following the authors prescription has a
discrepancy up to 25\% with respect to the authors result.
We also show that the attractive electrostatic force only due to the
surface potential differences was underestimated by a factor of 590.
The resulting disagreement with the experimental data might be
partially decreased by the effect of the charge double layer 
screening  which was not taken into account. 
All this leads to the conclusion that the results of this experiment are 
in fact uncertain.

\end{abstract}
\maketitle

In Ref.~\cite{1} the measurements of the attractive Casimir force
between an Au-coated sphere and a plate immersed in ethanol using
an atomic force microscope are presented. The obtained
experimental data are compared with the Lifshitz theory  taking
into account the frequency dependence of the dielectric functions
of Au and ethanol, and the correction due to surface roughness.
Consistency of the obtained data with Lifshitz's theory is
claimed, although at separation distances 
below 50\,nm, disagreement with theory has been observed which
increases with the decrease of separation. The performed experiment
is interesting as a test of the Lifshitz theory for applications to
three-layer systems and as a step towards the observation of
the Casimir repulsion predicted in such systems. However, as we
show below, the theoretical computations of the Casimir force
between smooth
Au surfaces separated by ethanol done according to the prescription 
 provided by Ref.~\cite{1}  leads to a discrepancy 
up to 25\% with  respect
to the results of Ref.~\cite{1}.
This
increases the observed discrepancy between experiment and theory. A 
second drawback is that the effect of the residual potential difference
between the sphere and the plate was calculated incorrectly and
significantly underestimated by a factor of 590. 
Finally, the interaction between the double layer 
formed in liquids, which would decrease the electrostatic force,
 was not taken into account without any justification.
%%%%%%%%%%%%%%%%%%%%%%%%%%%%%%%%%%%%%%%%%%%%%%%%%%%%%%%%%%%%%%
We show that the resulting electrostatic force is  of the same
order of magnitude as the Casimir force to be measured.
%%%%%%%%%%%%%%%%%%%%%%%%%%%%%%%%%%%%%%%%%%%%%%%%%%%%%%%%%%%%%
All these make the interpretation of this experiment uncertain.

The sphere-plate Casimir attraction across ethanol was
calculated using the Lifshitz formula (2) of Ref.~\cite{1}.
The dielectric function of Au along the imaginary frequency axis
was found using the Kramers-Kronig relation (Eq.~(3) in \cite{1})
and optical data from the optical properties
handbook \cite{2}. The dielectric function of
ethanol was taken from \cite{3} (it was previously used in
\cite{4}). The computational results are presented in Fig.~3
of Ref.~\cite{1} as a dotted line for perfectly smooth
surfaces of a sphere and a plate.

We have repeated the same computation using the same algorithm for
the determination of the dielectric function of Au along the
imaginary frequency axis, and the dielectric function of ethanol,
as it is explicitly presented in \cite{4}. Our results are shown
by the solid line in Fig.~1. In the same figure the computational
results obtained in Ref.~\cite{1} are presented as dots. 
As is seen in Fig.~1, the solid line significantly deviates from 
the dots over
the measurement range from $d=34\,$nm to $d=90\,$nm.
At $d=34$ and 40\,nm this deviation reaches 25\% and 21\%,
respectively, of the force magnitude shown by the dots.
It decreases to 6.5\% at $d=90\,$nm (at larger separations it is
impossible to retrieve the force values in Fig.~3 of Ref.~\cite{1}).

Our precise computation of the dielectric function of Au along the 
imaginary frequency axis on the basis of the
optical handbook data \cite{2} can be
found in Refs.~\cite{5,6}. The computations of the Casimir force
between Au surfaces using this function have been compared with
independent computations made by other groups \cite{7,8},  and
agreement at the level from 0.1\% to 0.2\% was achieved. 
Bearing in mind that the dielectric function of ethanol is
defined analytically \cite{4}, there is no problem in the
inclusion of an intermediate liquid layer. This permits us to
conclude that the dotted line in Fig.~1 is in disagreement with
the practically coincident computational results of three
above mentioned independent groups shown as the solid line.
As is claimed in \cite{1} with a reference to \cite{8a},
the calculated force between gold surfaces in vacuum can vary by 
as much as 5\% due to the variation in the optical properties
that occur for different samples. This point is however
irrelevant to the difference between the solid and dashed lines 
in Fig.~1, because if the data set reportedly used in Ref.~\cite{1}
is employed (tabulated data \cite{2} for Au and analytic function 
for ethanol), the results must be in agreement.

We admit that the roughness correction in Ref.~\cite{1} is
computed correctly (the roughness profiles are not provided
in \cite{1} and, thus, the respective computations cannot be
independently repeated), we arrive at the conclusion that the 
disagreement between experiment and theory in Ref.~\cite{1} is
larger than indicated. For example, at $d=40\,$nm the deviation
between experiment and theory in \cite{1} including surface
roughness is equal to $\approx 63$\% of the measured force
magnitude. If one takes into account, however, that the force
magnitude between smooth surfaces at $d=40\,$nm is  28.5\,pN
larger than indicated in \cite{1} (i.e., 163.5\,pN instead of
approximately 135\,pN) the deviation between experiment and theory
reaches $\approx 76$\%.

We emphasize that the computational results in Fig.~1, indicated
as dots, can be reproduced if one uses at all imaginary frequencies
the Drude dielectric function for both sphere and plate
materials,
\begin{equation}
\varepsilon_1=\varepsilon_2=1+\frac{\omega_p^2}{\xi(\xi+\gamma)},
\label{eq1}
\end{equation}
\noindent
instead of the optical handbook data, where $\omega_p=9.0\,$eV is the plasma
frequency and $\gamma=0.035\,$eV is the relaxation frequency.
It has been known \cite{7} that the use of the Drude function
instead of the optical tabulated data of \cite{2} leads to large
underestimation of the force magnitude at short separations.
Recall that  the text of Ref.~\cite{1} claims that the
Kramers-Kronig relation and the dielectric function of Au from
the handbook \cite{2} were used. The Drude dielectric function
is not mentioned in Ref.~\cite{1}.

The next point of this Comment is the effect of the residual potential
difference between the sphere and the plate. According to Ref.~\cite{1},
``the contact potential between the plate and the sphere in air is
$V_0=130\,$mV, as determined by varying the bias voltage on the
sphere while keeping the plate grounded.'' Reference \cite{1}
calculates the resulting electrostatic force in ethanol using the formula
\begin{equation}
F_{electrostatic}=-\frac{\pi R\varepsilon_0V_0^2}{\varepsilon_{ethanol}d}
=-10.326\,\mbox{pN{\ }at{\ }}d=40\,\mbox{nm}
\label{eq2}
\end{equation}
\noindent
and considers the obtained value negligible compared to the
Casimir force equal to --260\,pN.

This calculation is in error.  
In the process of measurements both the sphere and the plate in 
Ref.~\cite{1} were kept grounded. 
Theoretically, for grounded bodies the potential
difference $V_0$, as determined in the air, remains the same in the
presence of ethanol. In this case, the ground plays the role of a voltage 
source and compensates the screening effect of the polarization of
ethanol. However, Eq.~(\ref{eq2}) used in Ref.~\cite{1} is derived under 
the condition that at a fixed separation $d$
the total charge $Q$ on each plate of the capacitor is fixed and
is the same in air as well as in the
dielectric fluid
(Ref.~\cite{9}, page 30, problem 1.7,a). It is not applicable
to the case of grounded bodies where the charge cannot be controlled.
In this
situation, contrary to the authors statement on p.2 of Ref.~\cite{1}, 
the electrostatic force between the sphere and the plate is not
reduced but rather increases  
when the intervening material is a dielectric and not vacuum (see, e.g.,
Ref.~\cite{9}, page 30, problem 1.7,b). In electrodynamics
the residual potential difference remains equal to $V_0$
whether or not there is an intervening material while the capacitance
in the presence of ethanol is increased by $\varepsilon_{ethanol}$. 
Using the proximity force approximation, this leads to
a different equation for the residual electrostatic force
between the sphere and the plate
\begin{eqnarray}
&&
F_{electrostatic}=2\pi RE(d)=
-\frac{\pi R\varepsilon_0\varepsilon_{ethanol}V_0^2}{d}
\nonumber \\
&&\phantom{F_{electrostatic}}
=-6097\,\mbox{pN{\ }at{\ }}d=40\,\mbox{nm}.
\label{eq3}
\end{eqnarray}
\noindent
Here, $E(d)=-CV_0^2/(2S)$ is the energy per unit area of a parallel-plate 
capacitor of area $S$
with fixed potential difference $V_0$ between the plates
and $C=\varepsilon_0\varepsilon_{ethanol}S/d$ is the capacitance
in the presence of ethanol.

The analysis of real experiments should be guided by the above theoretical
considerations. Experimentally, there is no direct control either on $Q$ or
$V_0$ in liquids. 
While the variations of $Q$ can be arbitrary, the variations of
$V_0$ can be considered as small for grounded surfaces. The value of
$V_0$ is in fact controlled by the difference in the chemical
potentials of the two metals. For this reason the electrostatic force
can be estimated at constant $V_0$ rather than at constant $Q$.
This is still an estimation. In fact it can happen that even $V_0$ can
have small variations both depending on the separation and on the
introduction of ethanol.
Thus, in the experiment under consideration the residual electric 
force due to $V_0$ is
in fact much larger than the Casimir force and must be
taken into account in the comparison of experiment with theory.

The seeming contradiction between the large residual electric force
(\ref{eq3}) and the relatively small measured forces, as reported in Fig.~3 
of Ref.~\cite{1}, might be explained by an 
important effect that was not taken into account in 
Ref.~\cite{1} without any justification. 
This is the effect of the double layer formed near
the surfaces of the plate and the sphere in ionic solution \cite{10,12}.
Just like in a plasma, the ions from the salt impurities 
(even parts per million are sufficient while parts per thousand in
ethanol are possible \cite{pi1,pi2}) will
screen the surfaces leading to a decrease of the  magnitude
of the electrostatic force. This screening has been
experimentally investigated in Refs.~\cite{12,11} for a tip of an
atomic force microscope above a plate immersed in different
liquids. 
The electrostatic force itself can be attractive (always for conductive
surfaces which are grounded, as in Ref.~\cite{1}) or repulsive (e.g., 
for two insulating surfaces). Thus, in either case the double layer
effect acts to decrease the strength of the
attraction or repulsion.
The total electrostatic force between the grounded surfaces including 
the effect of the double layer in its simplest approximation takes the form 
(for a general case see Refs.~\cite{10,12})
\begin{equation}
F_{electrostatic}=-Ae^{-d/\lambda},
\label{eq4}
\end{equation}
\noindent
where $\lambda$ is Debye length and $A$ is a coefficient
depending on the parameters of the system. For a large Au sphere
of about $20\,\mu$m radius and a plate immersed in ethanol
the corresponding $A$ can be of several hundred pN. This is achieved
for ion concentration between $10^{-4}-10^{-5}$ moles per liter
for ethanol used in the experiment of Ref.~\cite{1}, 
much less than the solubility
limit for a typical impurity such as NaCl.

It is easily seen that the electrostatic force (\ref{eq4}) with 
account of the double layer is of the same order of magnitude as the
expected Casimir force and may contribute significantly to the
observed experimental data shown as circles in Fig.~3 of
Ref.~\cite{1}. An accurate account of
the effect of the net electrostatic force for the
experimental configuration used is necessary
before any comparison between experiment and theory can be made.

To conclude, Ref.~\cite{1} contains an interesting attempt to measure
the Casimir force in a three-layer system. The obtained results, however, 
cannot be considered as a ``precision measurement'' consistent
with Lifshitz's theory, as is claimed by the authors. In the above, we
have demonstrated large errors in the calculation of the Casimir force 
and, most notably, of the residual electrostatic force. We have also
indicated that the important effect of Debye screening 
due to the double layer which could partially compensate the large residual
electrostatic force was not taken into account. 
In our opinion, this effect is a realistic possibility to explain 
the large
disagreement between the experimental data in \cite{1} and the
magnitudes of the residual electrostatic forces between the grounded surfaces
calculated on the basis of classical electrodynamics.
The above remarks demonstrate that the results
of Ref.~\cite{1} concerning the observation of the Casimir force
in a three-layer system are uncertain. \hfill \\

This work was supported by the NSF Grant No.~PHY0653657 
(calculation of the Casimir force across a liquid) and
DOE Grant No.~DE-FG02-04ER46131 (calculation of the residual electric
forces).
G.L.K. and V.M.M. were also partially supported by
Deutsche Forschungsgemeinschaft, Grant No.~436\,RUS\,113/789/0--3.

%%%%%%%%%%%%%%%%%%%%%%%%%%%%%%%%%

%%%%%%%%%%%%%%%%%%%%%%%%%%%%%%%%
%%%__FIGURES__%%%%%%%%%%
%%%%%%%%%%%%%%
\begin{figure}
\vspace*{-4cm}
\centerline{
\includegraphics{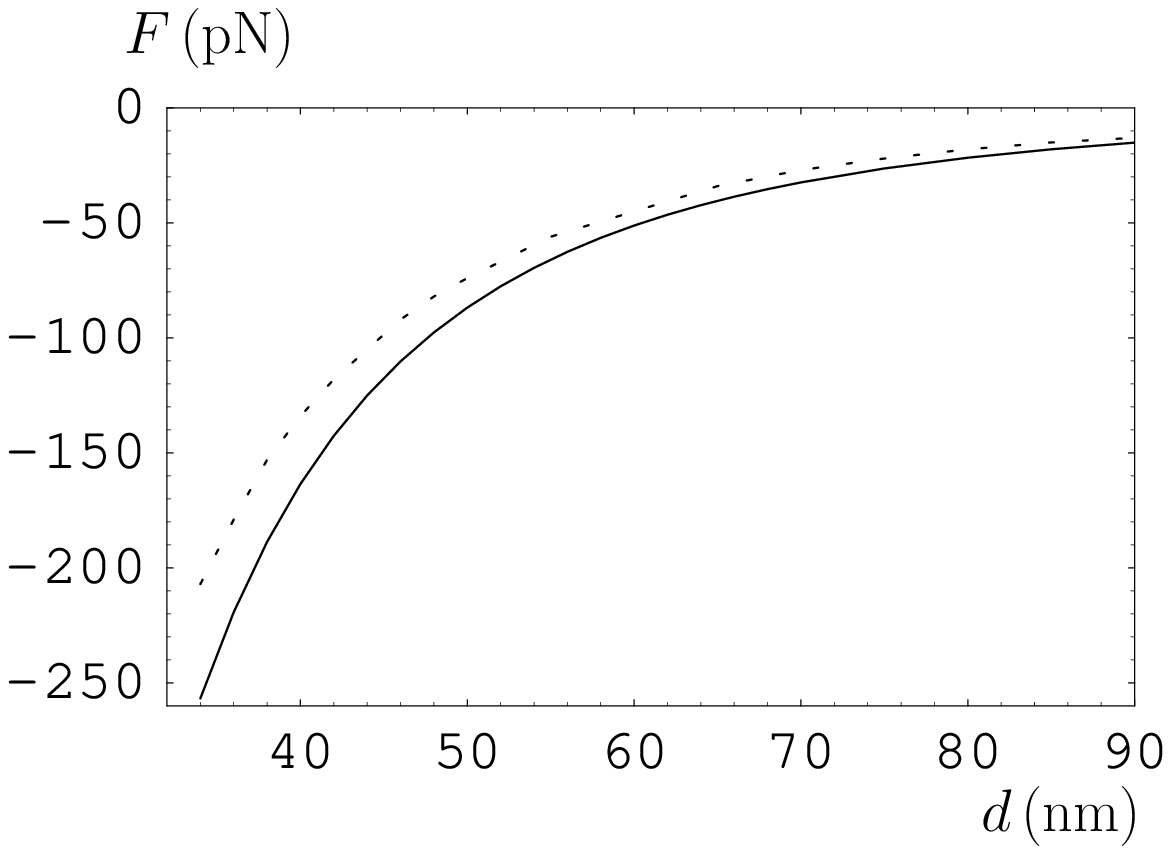}
}
\vspace*{-10cm}
\caption{
The Casimir force between perfectly smooth gold surfaces of
a sphere and a plate immersed in ethanol calculated in
Ref.~\cite{1} (dotted line) and in this Comment (solid line)
using the same procedure as a function of separation  (see
text for details).
}
\end{figure}

%%%%%%%%%%%%%%%%%%%%%%%%%%%%%%%%%%%%%%%%%%%%%%
\end{document}